# Modeling and Resource Allocation for HD Videos over WiMAX Broadband Wireless Networks

*Abdel Karim Al Tamimi, Chakchai So-In, and Raj Jain (IEEE Fellow), Washington University in St. Louis, St. Louis, MO*
*{aa7,cs5,jain}@cse.wustl.edu*

## 1. Introduction

Mobile video is considered a major upcoming application and revenue generator for broadband wireless networks like WiMAX and LTE. Therefore, it is important to design a proper resource allocation scheme for mobile video, since video traffic is both throughput consuming and delay sensitive.

In order to compare resource allocation schemes for mobile video, it is necessary to have an accurate model of the video traffic that represents real mobile videos. For limited-resource networks like WiMAX, it is essential to maximize the resources utilization. An accurate video model can provide the basis for a reliable traffic predictor that is the core component of any dynamic resource allocation scheme.

MPEG video frames are known to have seasonal characteristics. As shown in the Figure 1, MPEG video frames are divided into three types: I, P, and B-frames. These frame types differ in their function and size. For example, I-frames are the largest in size, and B-frames are the smallest in size.

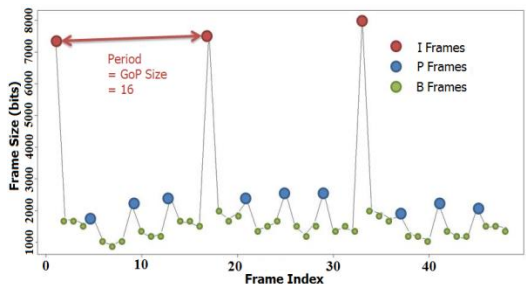

**Fig.1: Seasonal Characteristics of MPEG Video**

Typically, the pattern of video frames is repeated every "$s$" frames, where $s$ is the Group of Picture (GoP) size. This observation justifies our approach to model MPEG videos as a time series.

## 2. The SAM Model

Our mathematical model is based on the seasonal autoregressive integrated moving average (SARIMA) models [1,2]. SARIMA models aim to achieve better modeling by identifying both non-seasonal and seasonal parts of the data series.

The process of modeling a data series includes two steps: identifying the model order, and then estimating the model coefficients. The first step requires human intervention due to the significant statistical requirements in determining the best model. The second step uses algorithms like maximum likelihood (ML) to estimate the model's coefficients. SARIMA can be described as follows:

$$SARIMA = (p,d,q) \times (P,D,Q)^s \qquad (1)$$

where $p$ is the order of the autoregressive (AR) part; $q$ is the order of the moving average (MA) part; $d$ is the order of the differencing part. The parameters $P$, $Q$, and $D$ are the corresponding seasonal order, respectively. The parameter $s$ denotes the seasonality of the series. Our proposed model: Simplified Seasonal ARIMA model (SAM), as a SARIMA model, can be written as follows:

$$SAM = (1,0,1) \times (1,1,1)^z \qquad (2)$$

where $z$ is the video seasonality, which in most cases is equal to the used GoP size. This simplification means that SAM does not require any human intervention, and needs only 4 coefficients to be estimated. These coefficients are: AR coefficient ($\varphi$), MA coefficient ($\theta$), seasonal AR or SAR coefficient ($\Phi_s$), and seasonal MA or SMA coefficient ($\Theta_s$). SAM can be written in its difference form as:

$$\begin{aligned}X_t =\ & X_{t-1} + \varphi X_{t-1} - \varphi X_{t-2} + \Phi_s X_{t-s} - \varphi \Phi_s X_{t-s-1} \\ & - \Phi_s X_{t-s-1} + \varphi \Phi_s X_{t-s-2} - \theta \varepsilon_{t-1} - \Theta_s \varepsilon_{t-s} \\ & + \theta \Theta_s \varepsilon_{t-s-1} + \varepsilon_t\end{aligned}$$
$$(3)$$

SAM provides a unified approach to model video traces encoded with different video codec standards, using different encoding settings [1-3]. In recent research results [1, 2, 7], it was shown that SAM is capable of capturing the statistical characteristics of video traces within 5% of the optimal model for these video traces.

Figure 2 shows the modeling results of Star Wars IV with an AVC-encoded video trace. The model has been tested against video traces encoded using three different encoding settings and standards: MPEG-Part2, MPEG4-Part10/AVC (H.264), and AVC's scalable extension to support temporal scalability (SVC-TS).

In addition, in [3], we demonstrated that SAM has a clear edge in modeling high definition (HD) video traces over both AR modeling, and the automatic SARIMA estimation algorithm proposed in [4], that implements a unified approach to specify the model's order using a step-wise procedure.

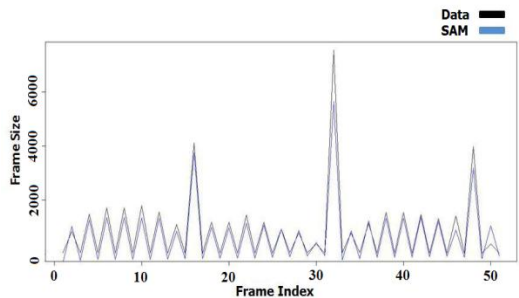
(a) Close-up Comparison

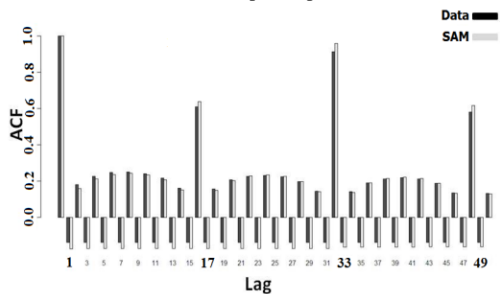
(b) ACF Comparison

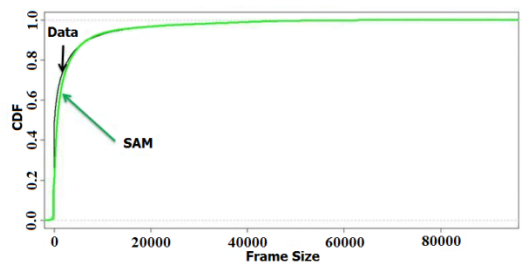
(c) CDF Comparison

**Fig. 2: SAM Modeling Results for Star Wars IV Movie Trace**

Based on our SAM model, we developed a trace generator that can be used to evaluate the performance of video based simulations. The trace generator provides the researchers with the ability to produce user-defined length traces that resemble the desired statistical attributes [2,5].

In [1], we presented the results of our analysis of various scheduling methods for Mobile WiMAX networks using the SAM traffic generator. These are Earliest Deadline First (EDF), Deficit Round Robin (DRR) and Easiest Deadline First with Deficit Round Robin (EDF-DRR). The simulation results for EDF, DRR, and EDF-DRR show that EDF is the most unfair. While EDF-DRR is an improvement, DRR is the most fair and provides the best performance for real-time mobile video traffic.

### 3. Video Traffic Prediction
SAM, as an accurate source model, can be used to predict future traffic based on the available short-term history of the incoming video frames. Using SAM difference's equation (3), future incoming video frames can be easily forecasted. We demonstrated in [6] that SAM provides 55% improvements on average, in terms of prediction accuracy compared to AR, and 53% improvements over the automatic SARIMA model estimation algorithm [4].

### 3. HD Video Traces Collection
We collected more than 50 HD video traces from the HD section of YouTube website that represents a wide variety of statistical characteristics. We encoded these traces using AVC codec with the most common settings, confirming experts' recommendations[6]. These traces provide the research community with the means to test and research new methods to optimize network resources. All the video traces are available to the research community through our website [5].

Our modeling and prediction comparisons are based on our HD video traces collection. The comparison results are available to the research community along with our developed tools [5].

In addition, we performed a full statistical analysis on our video traces collection. Our analysis included factor analysis using principle component analysis (PCA), and cluster analysis using *k-means* clustering [3].

### 4. Conclusions
The SAM model provides a convenient and accurate approach to model, generate and predict video traffic. It may be considered for practical solutions to solve the dynamic resource allocation challenge for live video streams, due to its ability to provide accurate results for the most common video codecs. This is especially important for networks with limited resources like WiMAX and LTE.

# IEEE COMSOC MMTC E-Letter

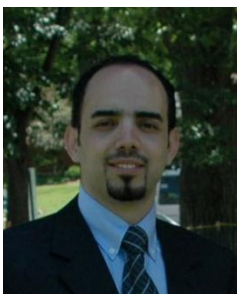

**Abdel Karim Al Tamimi** received his Bachelor's degree in Computer Engineering from Yarmouk University, Jordan in 2004. He received his Master's degree from Washington University in St. Louis in 2007. He is a final year PhD candidate in Computer Engineering at Washington Univerisity at St. Louis. His research interests include network systems, multimedia modeling, dynamic resource allocation, traffic engineering, and wireless networks.

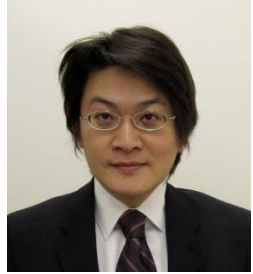

**Chakchai So-In** received the B.Eng. and M.Eng. degrees in computer engineering from Kasetsart University, Bangkok, Thailand in 1999 and 2001 respectively. He is currently working toward a Ph.D. degree at the Department of Computer Science and Engineering, Washington University in St. Louis, MO, USA. His research interests include architectures for future wireless networks/next generation wireless networks; congestion control in high speed networks; protocols to support network and transport mobility, multihoming, and privacy; and quality of service in broadband wireless access networks.

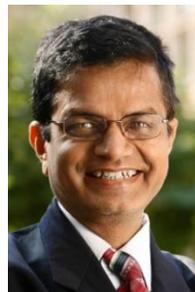

**Raj Jain** is a Fellow of IEEE, a Fellow of ACM, a winner of ACM SIGCOMM Test of Time award, CDAC-ACCS Foundation Award 2009, and ranks among the top 50 in Citeseer's list of Most Cited Authors in Computer Science. Dr. Jain is currently a Professor of Computer Science and Engineering at Washington University in St. Louis. He is the author of "Art of Computer Systems Performance Analysis," which won the 1991 "Best-Advanced How-to Book, Systems" award from Computer Press Association. His fourth book entitled " High-Performance TCP/IP: Concepts, Issues, and Solutions," was published by Prentice Hall in November 2003. He has co-edited "Quality of Service Architectures for Wireless Networks: Performance Metrics and Management," published in March 2010.